\documentstyle[12pt]{article}
\def\be{\begin{equation}}
\def\ee{\end{equation}}
\def\ba{\begin{array}}
\def\ea{\end{array}}
\input psfig

\begin{document}
\title{What is inside the nucleon?}
\author{ Manmohan Gupta and Harleen Dahiya    \\
{\it Department of Physics,} \\
{\it Centre of Advanced Study in Physics,} \\
{\it Panjab University,Chandigarh-160 014, India.}}   
 \maketitle

\begin{abstract}
We briefly review the structure of nucleon in the context of QCD,
Constituent Quark Model and Chiral Quark Model.
\end{abstract}

\section{Introduction}

The quest to peep into the successive layers of structure of
matter has led us from molecules to atoms 
and from atoms to subatomic particles and so on.
This quest, after 
painstaking efforts by experimentalists and theoreticians
together, in the present context has yielded a coherent
understanding of matter at the level of $10^{-18}$m. At
the present stage of scrutiny, the fundamental constituents
of matter are quarks and leptons interacting through gauge
bosons. There are six quarks, set up in well separated
three generations, for example,
\be
{\rm Quarks}: ~~~
\left( \begin{array}{c} u \\ d \end{array} \right)~~~
\left( \begin{array}{c} c \\ s \end{array} \right)~~~
\left( \begin{array}{c} t \\ b \end{array} \right).
\ee
This pattern is repeated for the leptons, each generation
containing a charged lepton and a corresponding neutrino,
for example,

\be
{\rm Leptons}: ~~~
\left( \begin{array}{c} e^- \\
\nu_e \end{array} \right)~~~
\left( \begin{array}{c} \mu^- \\
\nu_{\mu} \end{array} \right)~~~
\left( \begin{array}{c} \tau^- \\
\nu_{\tau} \end{array} \right).
\ee

At the present level of our understanding, quarks and
leptons are structureless objects having definite quantum
numbers and members of both categories carry spin half.
All matter, have to be made up of  quarks, leptons and
the corresponding antiparticles. Unlike leptons, quarks
have been ``seen'' only inside the hadrons. The theory
describing the interaction of these fundamental particles
is the $SU(3)_C \times SU(2)_L \times U(1)_Y$ gauge theory
called the minimal Standard Model (MSM). The MSM has two
distinct parts: `Quantum Chromodynamics' (QCD)
\cite{{frit},{koller},{qcd}}, described by the gauge group
$SU(3)_C$ and `Electroweak Model' \cite{{glas},{wei},{salam}}
described by
$SU(2)_L \times U(1)_Y$. The electroweak model describes
all possible electromagnetic and weak processes, with all
the electromagnetic interactions mediated by photons and the
weak processes, on the other hand, mediated by three massive
vector bosons, two charged ($W^{\pm}$) and one neutral
($Z^o$). The basic electromagnetic interaction is
characterized by the vertex given in Figure 1 where a charged
fermion couples to electromagnetic interaction. Interestingly,
all other e.m. interactions can be built from this basic
interaction. The weak interactions in the Standard Model are
characterized by emission and absorption of $W^{\pm}$ and
$Z^o$. In  Figure 2(a) we have shown the decay
$n \rightarrow p+e+{\bar {\nu}_e}$, mediated by charged
vector boson, whereas in Figure 2(b) we have shown the weak
interactions mediated by $Z^o$, usually called neutral current
interactions.

\begin{figure}
   \centerline{\psfig{figure=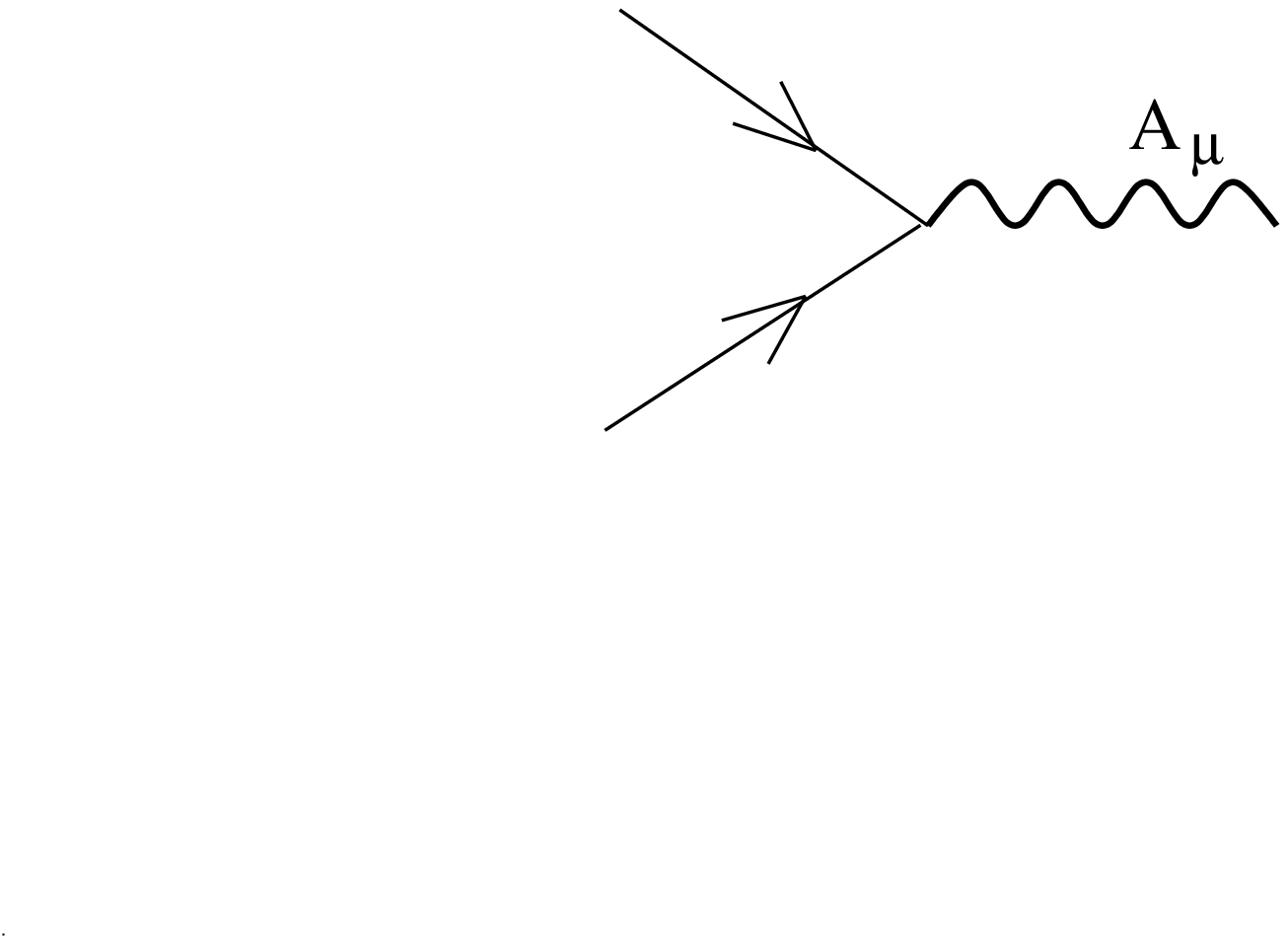,width=8cm,height=5cm}}
   \caption{Electromagnetic interaction.}
  \end{figure}

\begin{figure}
   \centerline{\psfig{figure=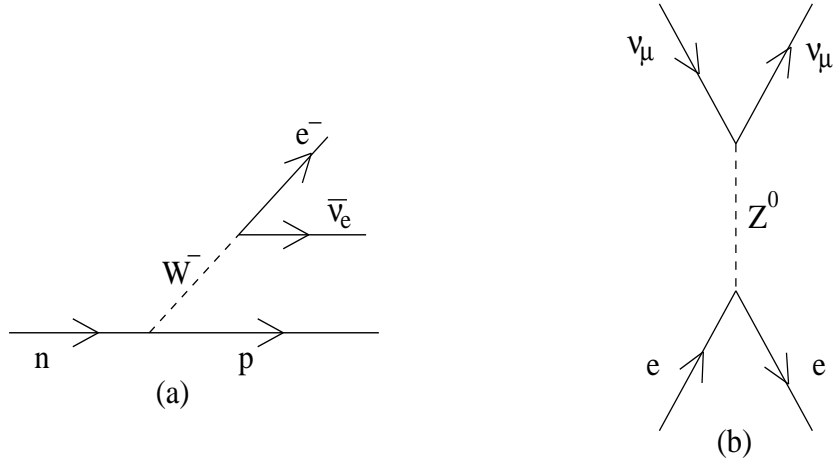,width=11cm,height=6cm}}
   \caption{Weak interaction.}
  \end{figure}

All hadrons (mesons and baryons) are made up of quarks and
antiquarks with $q-q$ and $q-\bar q$ interactions mediated
through gluons, the theory describing the interactions is
called Quantum Chromodynamics (QCD).
Naively speaking baryons are made up of three valence quarks
and mesons are made up of $q-\bar q$ combination. A proton,
for example, is made up of two $u$ quarks (each having $+2/3
|e|$ charge) and a $d$ quark having charge $-1/3 |e|$. Similarly
a neutron would consist of two $d$ quarks and a $u$ quark. In
Table 1, we have given the valence quark content of some of
the well known baryons and mesons. Since the basic purpose
of the article is to explore the structure of nucleon,
we, therefore, in the sequel detail some of the essentials
of QCD.

\begin{table}
\begin{center}
\begin{tabular}{|c|c|c|c|}       \hline              
Mesons & Quark content & Baryons & Quark content \\ \hline

$\pi^{\pm}$ & $u \bar d$, $d \bar u$ & p & uud \\
$\pi^o$ & $(u \bar u, d \bar d)$ & n & udd \\
$K^{\pm}$ & $u \bar s, s \bar u$ & $\Sigma^+$ & uus \\
$K^o, \bar K^o$ & $d \bar s, s \bar d$ & $\Sigma^o$ & dds \\
$\eta$ & $(u \bar u, d \bar d, s \bar s)$ & $\Xi^o$ 
& uss \\
& & $\Xi^-$ & dss \\
& & $\Lambda$ & uds \\ \hline
\end{tabular}
\caption{Valence quark structure of some of the important
mesons and baryons.}
\end{center}
\end{table}

\section{Quantum Chromodynamics (QCD)}
                                  
In QCD, quarks are 
endowed with an additional `color' degree of freedom,
for example, `red' (R), `green' (G) and `blue' (B)
with the  $q-q$ and
$q-\bar q$ interactions mediated by octet of colored gluons.
Gluons are similar to photons in that they have zero rest mass
and spin 1. However, photons carry no charge whereas gluons
carry color charge. The octet of colored gluons can be
characterised as follows:
\[  R \bar B~~~~ R \bar G~~~~ G \bar B~~~
\frac{1}{\sqrt 2}(B \bar B-G \bar G)  \]
\[  B \bar R~~~ G \bar R~~~ B \bar G~~~
\frac{1}{\sqrt 6}(B \bar B+G \bar G-2 R \bar R)  \]
Hadrons are colorless implying thereby that they are
color-singlet in the color space. Although the purpose of
the article is not to go into the technical details of the
gluon mediated $q-q$ interactions, however, in order
to facilitate the understanding of certain concepts, we would
mention the QCD  Lagrangian \cite{pro}, for example,  
\be
 L_{QCD}=-\frac{1}{4}F^{(a)}_{\mu \nu} F^{(a) \mu \nu} +
i\sum_{q} \bar{\psi}^i_q \gamma^{\mu} (D_{\mu})_{ij}
\psi^j_q- \sum_{q} m_q \bar{\psi}_q^i \psi_{qi}, \label{disl}
\ee
\be
 F^{(a)}_{\mu \nu}= \partial_{\mu}A^a_{\nu}-
 \partial_{nu}A_{\mu}^a+g_s f_{abc} A_{\mu}^b
A_{\nu}^c, 
\ee
\be
 (D_{\mu})_{ij}=\delta_{ij} \partial_{\mu}-i g_s \sum_{a}
 \frac{\lambda_{i,j}^a}{2} A_{\mu}^a, 
\ee

where $g_s$ is the QCD coupling constant and $f_{abc}$ are
the structure constants of the $SU(3)$ algebra, $\psi$ is the
quark field and $A$ is the gauge field.  The $q-q$ $q-\bar q$
interactions are usually discussed in terms of $\alpha_s$
which is related to the  QCD gauge coupling, $g_s$ as
\be
\alpha_s=\frac{g_s^2}{4 \pi}.
\ee
There are certain features of QCD which are quite distinctive
compared to Quantum Electrodynamics(QED). As is evident from
the definition of $F^{(a) \mu \nu}$, in terms of gluon field
$A^{(a)}_{\mu}$, there is a self interaction represented by
$g_sf_{abc} A^{(b)}_{\mu} A^{(c)}_{\nu}$ which is absent in
the case of photon mediated QED. The self interaction of
gluon in fact is a general property of any of the non-Abelian
gauge field theory. Another extremely important property of
QCD, which is also a general characteristic of the gauge
field theories, is the momentum dependence of the coupling
constants. The effective QCD coupling, $\alpha_s(Q^2)$,
can be shown to have the following momentum dependence at
momentum scale $Q$
\be
\alpha_s(Q^2)=\alpha_s(\mu^2)-{\alpha_s}^2(\mu^2)
\beta_o {\rm ln}(Q^2/\mu^2)+ ........
\ee
 where the $\beta_o$ is calculated to be

\be
 \beta_o=\frac{11 N_c-2 n_f}{12 \pi}.
\ee
 $N_c$ is the number of colors (=3), $n_f$ is the
number of active flavors, $i.e.$ the number of flavors
whose mass threshold is below the momentum scale, $Q$.
The corresponding momentum dependence of the  electromagnetic
fine structure constant, in the case of QED, is given as
\be
\alpha(Q^2)=\alpha(m^2)-\frac{{\alpha}^2(m^2)}{3 \pi}
{\rm log}(-Q^2/m^2)+ ........
\ee
 From the above expression, one can find out that
 $\alpha$ has the value 1/137
at energies which are not large compared with the electron
mass, however at LEP energies (101 GeV), it takes a value
closer to 1/128. In contrast to the electromagnetic
interactions, which is an Abelian gauge theory, the coupling
constant in the case of non-Abelian gauge theory decreases as
the energy increases, as can be checked from Equation (7).
Therefore, in the case of QCD, the coupling of one quark to
another is weak at very short distances or large four
momentum transfer squared ($Q^2$). The property
$\alpha_s \rightarrow 0$ as $Q \rightarrow \infty$ is
known as ``asymptotic freedom'' and  helps to explain why
quarks deep within hadrons are essentially free particles.
Conversely, the effective couplings grow as we go to large
distances and tends to infinity at very large distances or
small $Q^2$ and as a consequence it is impossible to separate
quarks. This property of increasing $\alpha_s$ at large
distances and consequently confinement of quarks is called
``infrared slavery''. At a deeper level the concept of
infrared slavery could be attributed to the self-interaction
of gluons in the case of QCD.

Naively speaking, thus we have two different pictures of
the inside of the nucleon, one at short distance and the
other at large distance. At sufficiently short distances,
which can be probed at sufficiently large energies, we can
consider quarks and gluons interacting very weakly with
each other, hence we can perform calculations of the
scattering cross-sections with the perturbative techniques.
At large distances, the nonlinearity of gluon-gluon couplings
come fully into play, as a consequence it is very difficult
to understand.

QCD has registered remarkable success in the limit
$Q^2 \rightarrow \infty$, where it is amenable to perturbative
calculations. In particular, it has helped in establishing
the quark degree of freedom, gluons \cite{{jade},{cello}} and
the presence of $q \bar q$ pairs inside the nucleon.
From the deep inelastic scattering, we learn that the
nucleon is composed of spin 1/2 point like particles which
can be identified with the quarks. However in the low energy
limit, in view of the intractability of the QCD, the
progress in this direction is painstakingly slow. However,
considerable insight has been achieved through lattice
calculations, QCD sum rules, $1/N_c$ expansions etc. The vast
amount of low energy data, however, is usually  explained
through Constituent Quark Model (CQM). In this context
CQM with QCD motivated spin-spin forces \cite{dgg} has been
extremely successful in explaining large amount of spin data.
This model, apart from providing considerable insight into the
hadronic matrix elements is devoid of complicated technicalities
from the calculation point of view.  For the benefit of the
reader, we include herewith a brief sketch of the CQM, so as to
enable one to understand simple calculations.

\section{The Constituent Quark Model with spin-spin forces}

The constituent quark model or naive quark model is based on
certain extremely simplifying assumptions. For example, hadrons
are made up of point like valence quarks, baryons consisting of
three quark combinations, whereas the mesons consisting of
quark-antiquark combinations. The valence quark content
of some of the baryons and mesons is given in Table 1. These
quarks interact through confining potential, several of these
have been used, the most popular being Coulombic + linear and
the harmonic oscillator. All hadronic transitions take place
through single quark transitions: for example, in a given
baryon,  two out of the three valence quarks would act as
spectators, whereas third quark will participate in the
interaction. Apart from the confining potential in the
CQM, some extra interactions between the quarks have also
been considered.

To understand the essentials of naive quark model, we discuss
in somewhat detail  a particular model, pioneered by
DGG \cite{dgg}, which has been extremely successful.
The starting point for CQM with chromodynamic spin-spin forces
is the Hamiltonian,

\be
H=L(\vec{r_1},\vec{r_2},....)+ \sum_{i}(m_i+
\frac{p_i^2}{2 m_i}+....) +\sum_{i>j} k \alpha_s S_{ij}.
\label{Ham}
\ee 
In the above equation, $L$ describes the universal interaction
responsible for quark binding, $\vec{r_i}, ~\vec{p_i}$ and
$m_i$ are the position, momentum and mass of the $i^{th}$ quark,
and $k$ is -4/3 for mesons and -2/3 for baryons. $S_{ij}$ is
the two-body Coulombic interaction and has the form:
\be
S_{ij}=\frac{1}{|\vec{r}|} \frac{1}{2 m_i m_j}
(\frac{\vec{p_i}.\vec{p_j}}{|\vec{r}|} + 
\frac{\vec{r_i}.(\vec{r}.\vec{p})}{{|\vec{r}|}^3} - 
\frac{\pi}{2} {\delta}^3(\vec{r}) (\frac{1}{m_i^2} +
\frac{1}{m_j^2} + \frac{16 \vec{S_i}.\vec{S_j}}{3 m_i m_j}) + .......  \label{Sij}
\ee
where $\vec{r}=\vec{r_i}-\vec{r_j}$ and $S_i$ is the spin of
the i$^{th}$ quark. The exact form of the confining potential
is not known, however, several kinds of confinement potential
have been used in the literature. To illustrate concrete
calculations of hadronic matrix elements we use harmonic
oscillator potential \cite{dgg} because of its exact
solvability and simplicity. The corresponding Hamiltonian
can be obtained from Equations (\ref{Ham}) and (\ref{Sij})
by replacing H by

\be
H=  \sum_{i}^{3}\frac{p_i^2}{2 m_i}+
\frac{1}{6} m \omega^{2} \sum_{i<j}{(r_i-r_j)}^2.
\ee

The spectrum of hadrons made up of $u$, $d$ and $s$ quarks, in the
CQM follows the SU(6)$\times$O(3) symmetry. According to
SU(6)$\times$O(3) symmetry, mesons fall into multiplets
$(35+1)$. The $35$ containing the nonet of scalar and vector
mesons. The baryons fall into the $(56 +{70}^{'} + {70}^{''}
+ 20)$ representations of $6 \times 6 \times 6$. The
extension to hadrons involving heavier quarks c, b and t,
can be carried out in the same manner. The O(3) symmetry
controls the spatial part of the wavefunction which
could be state of definite angular momentum and radial
excitations while constructing explicit wavefunctions of CQM.
The baryon wavefunction in CQM can be written as
\be
\psi_{{\rm Baryon}}=\phi_{{\rm unitary ~spin}}~
\chi_{{\rm spin}}~ \eta_{{\rm color}}
\ee
As the quarks are fermion spin 1/2 objects then the total wave
function has to be antisymmetric. The wave functions are
constructed in such a manner that the antisymmetricity
resides in the color space.
Baryons and mesons are colorless objects  which one ensures by
considering them to be color singlets. In the case of mesons
this is ensured by the $q-\bar q$ combinations, whereas in the
case of baryons this is ensured by taking the wave function
completely antisymmetric in color space.
We discard the color part of the wavefunction for
rest of our discussion as it does not play any dynamical
role in the low energy hadronic matrix elements. As an
explicit example of the symmetrized wave function in unitary
and spin space, we consider the ground state
octet of baryons, for which the wave function is expressed as

\be
\psi_o(8,{\frac{1}{2}}^{+}) = \frac{1}{\sqrt 2}.
(\phi^{'} \chi^{'} +\phi^{''} \chi^{''})\psi^s_o,
\label{8,1/2}
\ee 
where $\phi$, $\chi$ and $\psi$ denote respectively the SU(3),
spin and space wave functions, with the various types of
symmetry under quark exchange. For the proton and neutron,
we mention the explicit form of $\phi^{'}$, $\chi^{'},
\phi^{''}$, $\chi^{''}$, for example,
\be
\chi^{'} =  \frac{1}{\sqrt 2}(\uparrow \downarrow \uparrow
    -\downarrow \uparrow \uparrow), ~~~
 \chi^{''}
    =  \frac{1}{\sqrt 6} (2\uparrow \uparrow \downarrow
  -\uparrow \downarrow \uparrow
  -\downarrow \uparrow \uparrow),
\ee
\be
 {\phi}^{'}_p = \frac{1}{\sqrt 2}(udu-duu),~~~
{\phi}^{''}_p = \frac{1}{\sqrt 6}(2uud-udu-duu),
\ee
\be
\phi^{'}_n = \frac{1}{\sqrt 2}(udd-dud),~~~
 \phi^{''}_n = \frac{1}{\sqrt 6}(udd+dud-2ddu).
\ee

The spatial wavefunctions, denoted by $\psi$, are solutions
of the Hamiltonian. Each level will have definite symmetry
and will be associated with the SU(6) to build a symmetric
function. The levels and their symmetries are dependent on
the potential, generally, the ground-state level is symmetric
with $L^P=0^+$, and the next level is of mixed symmetry with
$L^P=1^-$. The wavefunctions of the first few spatial states
are listed as under. The ground state is given by
\be
\psi^s(56,0^+)=\psi_0(\rho,\lambda),
\ee
the N=1 states are as

\be
\psi^{'}(70,1^{-})={(\frac{8}{3} \pi)}^{1/2} R^{-1} Y^M_1(\rho) 
\psi_0(\rho,\lambda),
\ee

\be
\psi^{'}(70,1^{-})={(\frac{8}{3} \pi)}^{1/2} R^{-1} Y^M_1(\lambda) 
\psi_0(\rho,\lambda),
\ee

where as the N=2 states are

\be
\psi^{s}(56,0^{+})={(\frac{1}{3})}^{1/2} R^{-2}
[3 R^2-(\rho^2+\lambda^2)] \psi_0(\rho,\lambda), \label{s}
\ee
\be
\psi^{'}(70,0^{+})={(\frac{1}{3})}^{1/2} R^{-2}[2 \lambda.\rho)]
\psi_0(\rho,\lambda),
\ee
\be
\psi^{''}(70,0^{+})={(\frac{1}{3})}^{1/2} R^{-2}[\rho^2-
\lambda^2] \psi_0(\rho,\lambda),         \label{''}
\ee
\be
\psi^{''}(70,2^{+})={(\frac{8}{3} \pi)}^{1/2} R^{-2}
[Y^M_2(\rho)-Y^M_2(\lambda)] \psi_0(\rho,\lambda).
\ee
For rest of  spatial wavefunctions we refer the reader to
reference \cite{yaouanc}.
The variables $\rho$ and $\lambda$ are defined as
\be
\rho =\frac{1}{\sqrt 2}(r_1-r_2), ~~
\lambda =\frac{1}{\sqrt 2}(r_1+r_2-2 r_3), ~~
R =\frac{1}{3}(r_1+r_2+r_3).
\ee
$\rho$ being antisymmetric in 1 and 2 and $\lambda$ being
symmetric in 1 and 2.

Spin-spin forces lead to interband mixing, for example, the
octet wavefunction $\psi (8,{\frac{1}{2}}^{+})$ not only gets
contribution from the octet $|56,0^+>_{N=0}$ but also 
from $|56,0^+>_{N=2}, |70,0^+>_{N=2}$ and
$|70,2^+>_{N=2}$. Therefore, the nucleon  wavefunction is
expressed as,

\clearpage

\[ \psi (8,{\frac{1}{2}}^{+})=|56,0^+>_{N=0}+\alpha |56,0^+>_{N=2}\]
\be
+ \beta |70,0^+>_{N=2}+ \epsilon |70,2^+>_{N=2}. \label{full}
\ee

Using the above Hamiltonian of DGG, Isgur and Collaborators
have carried an  extremely detailed analysis of hadronic
spectra and hadronic matrix elements. For detailed exposure
in this regard we refer the reader to references
\cite{{dgg},{yaouanc},{mgupta1},{Isgur1}}. For the special
case of nucleon they arrive at the following wave function
                          
\[ {\left|8,{\frac{1}{2}}^+ \right>}_n = 0.90 |56,0^+>_{N=0}
-0.34 |56,0^+>_{N=2} \]
\be
-0.27 |70,0^+>_{N=2} - 0.06 |70,2^+>_{N=2}.
\ee

In Equation(\ref{full}) it should be noted that
$(56,0^+)_{N=2}$ does not affect the
spin-isospin structure of  $(56,0^+)_{N=0}$, whereas
$(70,0^+)_{N=2}$ does not affect the spin-isospin structure
of $(70,2^+)_{N=2}$. Therefore, Equation(\ref{full}) can be
simplified to

\begin{equation}
\left|8,{\frac{1}{2}}^+ \right> = {\rm cos} \phi |56,0^+>_{N=0}
+ {\rm sin} \phi|70,0^+>_{N=2}, \label{mixed}
\end{equation}
with $\phi=20^o$. This has been referred to as non-trivial mixing
in the literature{\cite{mgupta1}.

This CQM with one gluon mediated $q-q$ and $q-\bar q$ forces
have been applied very successfully to large variety of low
energy hadronic matrix elements \cite{manohar}. It has not
only given a remarkably accurate description of hadron
spectroscopy data \cite{Isgur1} but has also been able to
describe some very subtle features of the data, such as
neutron charge radius \cite{{yaouanc},{em}}, $N-\Delta$ mass
difference, photohelicty amplitudes \cite{photo},
baryon magnetic moments, etc.

To illustrate the success of CQM, in the sequel we discuss
two cases, for example, nucleon magnetic moments and neutron
charge radius. The magnetic moment for a nucleon is defined as
\be
\mu(B)=\Delta u^{B} \mu_u+\Delta d^{B} \mu_d+\Delta s^{B} \mu_s,
\label{mag}
\ee
where $\Delta u^{B}, \Delta d^{B}$ and  $\Delta s^{B}$
for the given nucleon are the spin polarizations defined as:
\be
\Delta q=q^{\uparrow}-q^{\downarrow},
\ee
$q^{\uparrow}$ and $q^{\downarrow}$ being the number of quarks
with spin up and spin down. The total spin
\be
\Delta \Sigma = \Delta u + \Delta d +\Delta s,
\ee
is twice the value of S (spin of proton).

Assuming $u$ and $d$ to be point particles, the magnetic moments
associated with these ($\mu_u, ~\mu_d$) can be defined as 
\be
\mu_u=\frac{q_u}{2 m_u}=\frac{2}{2 m_N}=2 \mu_N,
\ee
\be
\mu_d=\frac{q_d}{2 m_d}=-\frac{1}{2 m_N}=-\mu_N.
\ee
Here the masses of $u$ and $d$ quarks are considered to be 1/3 rd of
the nucleon mass and are taken to be equal.
To find the number of quarks with spin up and spin down,
$q^{\uparrow}$ and $q^{\downarrow}$, say in proton, one has to
consider the symmetrized wavefunction for the proton given in
Equation (\ref{8,1/2}). Let us calculate the number of u
quarks with spin up. This can be
calculated by considering the expectation value of the operator
$n_u(i) P_{\uparrow}(i)$, where i stands for the i$^{th}$ quark
and $P_{\uparrow}(i)$ is the projection operator for spin up and
is 1 if the i$^{th}$ quark has spin up and 0 otherwise. $n_u(i)$
is 1 if the i$^{th}$ quark is $u$ and 0 otherwise. Then,
\be
u^{\uparrow}= <56, 0^+|\sum_{i=1}^{3} n_u(i) P_{\uparrow}(i)
|56, 0^+>.
\ee
Since the wavefunction in Equation(\ref{8,1/2}) is symmetrized,
as well as  the operator $n_u(i) P_{\uparrow}(i)$ does not
affect the spatial part of the wavefunction, therefore one
can write
\[ u^{\uparrow}=\frac{3}{2}<\chi^{'} \phi^{'} +
\chi^{''} \phi^{''} |n_u(3) P_{\uparrow}(3)|\chi^{'} \phi^{'} +
\chi^{''} \phi^{''}>. \]
By carrying out simple calculation one finds
$u^{\uparrow}=\frac{5}{3}$. Similarly we can find
$u^{\downarrow}, d^{\uparrow}$ and $d^{\downarrow}$,
for example,
\be
u^{\downarrow}=\frac{1}{3}, ~~
d^{\uparrow}=\frac{1}{3}, ~~ d^{\downarrow}=\frac{2}{3}.
\ee
This can be repeated for other baryons, for example, in
neutron we interchange $u$ and $d$, in $\Sigma^+$ we replace
$d$ by $s$ and so on.
 
Thus, the contribution by each of the quark flavors to the 
proton spin can be written as:
\be 
\Delta u=\frac{4}{3}, ~~\Delta d=-\frac{1}{3},~~
\Delta s=0. \label{56}
\ee 
From Equations (\ref{mag}) and (\ref{56}) we get the
magnetic moment of the proton and the neutron as,
\be
\mu(p)=\frac{4}{3} \mu_u-\frac{1}{3} \mu_d, \label{mup} 
\ee
\be
\mu(n)=\frac{4}{3} \mu_d-\frac{1}{3} \mu_u. \label{mun}
\ee
If we substitute $\mu_u=-2 \mu_d$, then we obtain 
$\frac{\mu_p}{\mu_n}=-\frac{3}{2}$ which is very well in
agreement with the experiment. The same thing can be repeated
when we use the wavefunction with non-trivial mixing. The
magnetic moments are
 
\be
\mu(p)={\rm cos}^2 \phi(\frac{4}{3} \mu_u-\frac{1}{3} \mu_d)+
{\rm sin}^2 \phi(\frac{2}{3} \mu_u+\frac{1}{3} \mu_d),
\ee
and
\be
\mu(n)={\rm cos}^2 \phi(\frac{4}{3} \mu_d-\frac{1}{3} \mu_u)+
{\rm sin}^2 \phi(\frac{2}{3} \mu_d+\frac{1}{3} \mu_u).
\ee
One can calculate magnetic moment of other baryons, and in
Table 2, we have presented the results of a particular
calculation \cite{mgupta1}.

\begin{table}
\begin{center}
\begin{tabular}{|c|c|c|c|} \hline
Baryons & Expt. value & CQM without & CQM with  \\ 
&&configuration mixing & configuration mixing \\ \hline
$\mu(p)$ &2.793 & 3             & 2.766 \\
$\mu(n)$ & -1.913 &-2            & -1.766 \\
$\mu(\Sigma^+)$ & 2.42 & 2.86       & 2.68 \\
$\mu(\Sigma^-)$ & -1.105 &  -1.13   & -1.084 \\
$\mu(\Xi^o)$ & -1.25 &   -1.46      &-1.43 \\
$\mu(\Xi^-)$ & -.69 &  -.46       &-.66 \\    \hline
\end{tabular}
\caption{Magnetic moments of baryons in CQM.}
\end{center}
\end{table}

There are large number of hadronic
matrix elements which can be calculated and these agree very
well with the data. For the sake of readability of the article
we include another calculation of CQM which involves calculations
of spatial wavefunctions. The neutron charge radius is usually
expressed in terms of the slope of the electric form factor
$ G_{E}^n(q^2)$, the experimental value {\cite{kopecki}} of
which is given as

 \be  (\frac{d G^n_E (q^2)}{d|q^2|})_{q^2=0} =
    0.47 \pm 0.01 ~GeV^{-2}.
 \ee

If we assign the nucleon to a pure 56 (with the spin expressed in
terms of Pauli spinors ), the neutron electric form factor
vanishes for all $q^2$.
Considering our complete wavefunction with the $56-70$ mixing and 
performing the calculations, keeping the lowest order in tan$\phi$,
we obtain
\be
\mu_n\frac{G^n_E (q^2)}{G^n_M (q^2)} = -{\rm tan}\phi {\sqrt 2}
<{\psi}^{''}_{N=1}|e^{i {\vec q}.{\vec r_3}}|{\psi}^s>,
\ee
and from Equations (\ref{s}) and (\ref{''}) we have
\be
<{\psi}^{''}_{N=1}|e^{i {\vec q}.{\vec r_3}}|{\psi}^s>=
\frac{|q^2| R^2}{6 \sqrt 3}.
\ee
Neutron charge radius can be expressed as

 \be
 <r_n^2>= 6 (\frac{d G^n_E (q^2)}{d|q^2|})_{q^2=0} =
\sqrt{\frac{2}{3}} R^2 (-{\rm tan} \phi),
 \ee
where $\phi$ is the mixing angle and is negative and $R^2$
is the shape factor {\cite{{yaouanc},{mgupta1}} for the
harmonic oscillator wave function. The calculated values
of neutron charge radius $<r^2_n> (=6b)$ as function of
$\phi$ and $R^2$ are presented in Table 3. From the table
one can immediately find out that CQM with spin spin forces
is able to give an excellent fit to neutron charge radius.

\begin{table}
\begin{center}
\begin{tabular}{|c|c|c|c|} \hline
& $\phi$ & $R^2 (GeV^{-2})$ 
 & $<r_n^2> (GeV^{-2})$  \\  \hline
Expt. value &  & - & 2.82 \\ \hline
  & -$20^0$  & 8 & 2.64 \\
              &          & 9 & 2.93 \\
                    &       & 10& 3.23 \\

Calculated        & -$18^0$ & 8 & 2.39 \\
values            &       & 9 & 2.65  \\
                  &       &10 & 2.91  \\

             & -$16^0$ & 10 & 2.61 \\
                      &       & 11 & 2.84 \\
                    &       & 12 & 3.07 \\      \hline

\end{tabular}
\caption{Neutron charge radius in CQM.}
\end{center}
\end{table}

\section{Difficulties with CQM}

Despite amazing success in explaning large and diverse
amount of hadronic data, the basic tenents of CQM raise
many questions about their justifications. Neither one can
deduce it from basic QCD considerations nor can one provide
justification for its basic assumptions. So,
from aesthetic considerations it is very unsatisfactory situation.

Besides the philosophical inadequacy of CQM, there are a few
parameters which have defied explanation within CQM. For example,
$G_A/G_V$, defined in terms of the spin distribution functions,
is given as
 \be
G_A/G_V=\Delta u-\Delta d.  \label{gav}
\ee
In comparison to the experimental value of 1.26, Equation
(\ref{gav}) predicts it to be $\frac{5}{3}$ in the case of CQM.
 Introduction of configuration mixing
does not help much in this case, for example, with a mixing
characterised by $\phi=20^o$, $G_A/G_V$ doesn't change much.
Similarly, there are several other
parameters which require one to go beyond CQM.  

The most important challenge to CQM was, however, posed by the
observations in the deep inelastic scattering of the polarised
leptons off polarised nucleons made by the
European Muon Collaboration (EMC) \cite{EMC}.
The deep inelastic polarized muon-proton scattering measurements
made by the EMC indicated  that the entire spin of the proton is
not carried by the valence quarks but only 30\%
of the spin is carried by the valence quarks. It also indicated
that the $q \bar q$ sea is not unpolarised and there is a  
significant contribution to the proton spin by the strange quarks
in the sea. This is contrasted with the CQM assumptions that
entire spin of the proton is carried by the valence quarks.
This was called ``spin crisis''.  

Further, in CQM, the similarity of the $u$ and $d$ quark masses and
the flavor independent nature of the gluon couplings led to
expect $\bar d=\bar u$, thus to the validity of the Gottfried
sum rule in CQM. The NMC measurements of the muon scatterings
off proton and neutron targets \cite{NMC}, however, show that
the Gottfried sum rule \cite{GSR} is violated. It has been
interpreted as showing $\bar d>\bar u$ in the
proton. This conclusion has been confirmed by NA51 \cite{na51}
in the Drell-Yan process with
proton and neutron targets.

From the above discussion it seems that many of the successes of
CQM are primarily due to cancellation which are taking place due
to various degrees of freedom inside the nucleon.
It therefore becomes interesting to introduce components into the
wavefunction having angular momentum, a polarized sea of
quark-antiquark pairs, gluons and Goldstone bosons etc.
In this context we would like to discuss a particular successful
model, Chiral Quark Model, which not only incorporates
the basic features of CQM but also some other degrees of freedom.

\section{Chiral Quark Model ($\chi$QM)}

Chiral quark model was developed \cite{{manohar},{wein}}
essentially to understand the successes of CQM.   
The idea of the $\chi$QM is based on the picture that a quark
inside a nucleon emits quark-antiquarks pairs via
Goldstone bosons (GB), for example,
\be
  q_{\pm} \rightarrow GB^{0}
  + q^{'}_{\mp} \rightarrow  (q \bar q^{'})
  +q_{\mp}^{'}.  \label{gb}
\ee

\begin{figure}
   \centerline{\psfig{figure=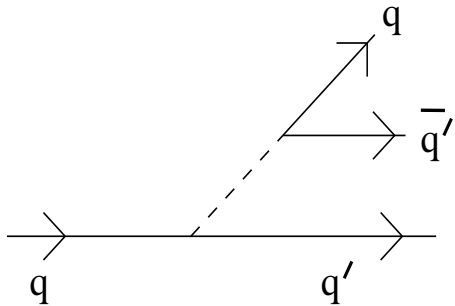,width=6cm,height=4cm}}
 \caption{Production of a $q-\bar q$ pair via a Goldstone
Boson emission.}  
  \end{figure}

The basic interaction causes a modification of the spin content
because a quark changes its helicity by emitting a spin zero meson.
It causes a modification of the flavor content because the GB
fluctuation, unlike gluon emission, is flavor dependent. The spin
flip process makes it possible to understand the spin content of the
nucleon, which was not possible in the conventional constituent
quark model. With the  admixture of mesons to the nucleon
wavefunction, one finds that only $\frac{1}{3}$rd of the nucleon is
carried by the quarks. Moreover, for the other spin-flavor
observables, such as magnetic moments, sea quark distributions and
the Gottfried sum rule, the agreement with experimental data is
also improved using this model. Thus, inside the nucleon, but not
deep inside where perturbative QCD is applicable, the
effective degrees of freedom are constituent quarks, gluons, the
$\chi SB$ Goldstone bosons and the $q-\bar q$ pairs.
In order to make the article self contained we reproduce in the
sequel some of the essential details of $\chi$QM.

The basic idea of $\chi$QM is that the chiral symmetry breaking
takes place at a distance significantly smaller than the
confinement scale. For example, the QCD confinement scale is
characterised by $\Lambda_{QCD}$=0.1-0.3 GeV,  whereas the
chiral symmetry breaking scale, $\Lambda_{\chi SB}$ is
characterised by 1 GeV.  The Lagrangian based on the
chiral quark model is
\be
 L = g_8 \bar q \phi q,  \label{chil}
\ee
where $g_8$ is the coupling constant,
\[ q =\left( \ba{c} u \\ d \\ s \ea \right),\]
and 
\be
\phi = \left( \ba{ccc} \frac{\pi^o}{\sqrt 2}
+\beta\frac{\eta}{\sqrt 6}+\zeta\frac{\eta^{'}}{\sqrt 3} & \pi^+
  & \alpha K^+   \\
\pi^- & -\frac{\pi^o}{\sqrt 2} +\beta \frac{\eta}{\sqrt 6}
+\zeta\frac{\eta^{'}}{\sqrt 3}  &  \alpha K^o  \\
 \alpha K^-  &  \alpha \bar{K}^o  &  -\beta \frac{2\eta}{\sqrt 6}
 +\zeta\frac{\eta^{'}}{\sqrt 3} \ea \right).
\ee

SU(3) symmetry breaking is introduced by considering
different quark masses $m_s > m_{u,d}$ as well as by considering
the masses of Goldstone Bosons to be non-degenerate
$(M_{K,\eta} > M_{\pi})$ {\cite{{cheng1},{song},{johan}}}, whereas 
the axial U(1) breaking is introduced by
$M_{\eta^{'}} > M_{K,\eta}$
{\cite{{{cheng1},{song},{johan},{cheng}}}.
The parameter $a(=|g_8|^2$) denotes the transition probability
of chiral fluctuation
of the splittings  $u(d) \rightarrow d(u) + \pi^{+(-)}$, whereas 
$\alpha^2 a$ denotes the probability of transition of
$u(d) \rightarrow s  + K^{-(0)}$.
Similarly $\beta^2 a$ and $\zeta^2 a$ denote the probability of
$u(d,s) \rightarrow u(d,s) + \eta$ and
$u(d,s) \rightarrow u(d,s) + \eta^{'}$, respectively.

The effective
Lagrangian in the region between $\Lambda_{\chi SB}$ and
$\Lambda_{QCD}$ has fundamental quark and gluon fields, because
these particles are not bound into color-singlet hadrons at such
short distances. Since the  $SU(3)_L \times SU(3)_R$ global chiral
symmetry is spontaneously broken, there is also an octet of
pseudoscalar Goldstone boson, which are put in as fundamental
fields. The  Goldstone boson fields are essential if the Lagrangian
is to consistently reproduce the effects of a spontaneously broken
global symmetry.
In this energy range the quarks and GBs propagate in the QCD
vacuum which is filled with the $q- \bar q$ condensate. The
interaction of a quark with the condensate will cause it to
gain an extra mass of $\simeq 350$ MeV. This is the $\chi$QM
explanation of the large constituent quark mass. The QCD
Lagrangian is also invariant under the axial U(1) symmetry,
which would imply the ninth GB $m_{\eta^{'}} \simeq m_{\eta}$.
But the existence of axial anomaly breaks the symmetry and
in this way the $\eta^{'}$ picks up an extra mass.

The interaction of the GBs is weak enough to be treated by
perturbation theory. This means that on long enough time
scales for the low energy parameters to develop we have

\[ u^{\uparrow} \rightleftharpoons (d^{\downarrow} + \pi^+) + 
(s^{\downarrow} + K^+)+(u^{\downarrow} + \pi^o,\eta,\eta^{'}),\]

\[ d^{\uparrow} \rightleftharpoons (u^{\downarrow} + \pi^-) + 
(s^{\downarrow} + K^o)+(d^{\downarrow} + \pi^o,\eta,\eta^{'}),\]

\[ s^{\uparrow} \rightleftharpoons (u^{\downarrow} + K^-) + 
(d^{\downarrow} + \bar K^{o})+(s^{\downarrow} +\eta, \eta^{'}).\]

In the absence of interactions, the proton is made up of two
$u$ quarks and one $d$ quark. Proton's flavor content can be
calculated after any one of these quarks turns into part of
the quark sea by `disintegrating', via GB emissions, into a
quark plus a quark-antiquark pair.

$\chi$QM with SU(3) symmetry ($\alpha= \beta =1$) is
also able to provide fairly satisfactory explanation for
various quark flavor contributions to the proton spin
{\cite{eichten}}, baryon magnetic moments
\cite{{cheng},{eichten}} as well as the absence of
polarizations of the antiquark sea  in the nucleon
{\cite{{song},{antiquark}}} . However, in the case of
hyperon decay parameters  the predictions of the
$\chi$QM are not in tune with the data   {\cite{decays}},
for example, in comparison to the experimental numbers
0.21 and 2.17 the $\chi$QM with SU(3) symmetry predicts
$f_3/f_8$ and $\Delta_3/\Delta_8$ to be $\frac{1}{3}$ and
$\frac{5}{3}$ respectively. It has been shown
{\cite{{song},{cheng}}} that when SU(3) breaking  effects are
taken into consideration within $\chi$QM, the  predictions of
the $\chi$QM regarding the above mentioned ratios
 have much better overlap with the data.

However, as mentioned earlier that the constituent quark model
with one gluon mediated configuration mixing (CQM$_{gcm}$) gives a
fairly satisfactory explanation of host of low energy hadronic
matrix elements {\cite{{dgg},{Isgur1},{em}}}.
In view of the fact that constituent quarks constitute one of
the important degrees of freedom, therefore it becomes interesting
to examine, within the $\chi$QM, the implications
of one gluon mediated configuration mixing. This is
particularly interesting as some of the low energy
data are responsive only to configuration mixing.

\section{Chiral Quark Model with configuration mixing}
In order to make the article self contained we discuss here
some of the essential details of the
chiral quark model with one gluon generated configuration mixing
($\chi$QM$_{gcm}$), for the
details we refer the reader to reference \cite{hd}.
To begin with, we consider the spin distribution functions
for nucleon wave function affected by spin-spin forces.
However, for the sake of simplicity we consider nucleon
wave function described by Equation (\ref{mixed}).
The spin distribution functions for proton are defined as
\[ \left< 8,{\frac{1}{2}}^+|N|8,{\frac{1}{2}}^+\right>={\rm cos}^2
\phi <56,0^+|N|56,0^+> \]
\be
+{\rm sin}^2 \phi<70,0^+|N|70,0^+>.
\ee
For the $|56>$ part we have
\be
 <56,0^+|N|56,0^+>=\frac{5}{3} u^{\uparrow} +\frac{1}{3}
 u^{\downarrow}+ \frac{1}{3} d^{\uparrow} +\frac{2}{3}
 d^{\downarrow},
\ee
as derived earlier. In the case of $|70>$,
one can find the spin in the similar manner and are defined as
\be
 <70,0^+|N|70,0^+>=\frac{4}{3} u^{\uparrow} +\frac{2}{3}
 u^{\downarrow}+ \frac{2}{3} d^{\uparrow} +\frac{1}{3}
 d^{\downarrow}.
\ee

In the  $\chi$QM, the basic process is the emission of a Goldstone
Boson which further splits into $q \bar q$ pair as mentioned in
Equation (\ref{gb}) of the
text. Following reference \cite{johan}, the spin structure
after one interaction can be obtained by  substituting for every
quark, for example,

\be
 q^{\uparrow} \rightarrow P_q q^{\uparrow} +
 |\psi(q^{\uparrow})|^2,
 \ee
where $P_q$ is the probability of no emission of GB from a $q$
quark and the probabilities of transforming a $q^{\uparrow
\downarrow}$ quark are $|\psi(q^{\uparrow})|^2$, given as

\be
 |\psi(u^{\uparrow})|^2=\frac{a}{6}(3+\beta^2+2 \zeta^2)
 u^{\downarrow}+ a d^{\downarrow}+a \alpha^2 s^{\downarrow},
\ee

\be
 |\psi(d^{\uparrow})|^2=a u^{\downarrow}+ 
\frac{a}{6}(3+\beta^2+2 \zeta^2)d^{\downarrow}+
a \alpha^2 s^{\downarrow},
\ee

\be
 |\psi(s^{\uparrow})|^2=   a \alpha^2 u^{\downarrow}+ 
a \alpha^2 d^{\downarrow}+\frac{a}{3}
(2 \beta^2+\zeta^2)s^{\downarrow}.
\ee
The quantity of interest here is $\hat B$, defined
using the above Equations

\[ \hat B={\rm cos}^2 \phi \left[ \frac{5}{3}(P_u u^{\uparrow} +
|\psi(u^{\uparrow})|^2)+
\frac{1}{3}(P_u u^{\downarrow} + |\psi(u^{\downarrow})|^2)+
\frac{1}{3}(P_d d^{\uparrow} + |\psi(d^{\uparrow})|^2)
\right. \]
\[ \left. +\frac{2}{3}(P_d d^{\downarrow} + |\psi(d^{\downarrow})|^2)
\right] +{\rm sin}^2 \phi \left[ \frac{4}{3}(P_u u^{\uparrow}
+|\psi(u^{\uparrow})|^2)+ \frac{2}{3}(P_u u^{\downarrow}
+|\psi(u^{\downarrow})|^2) \right. \]
\be
\left.
+\frac{2}{3}(P_d d^{\uparrow} + |\psi(d^{\uparrow})|^2)+
\frac{1}{3}(P_d d^{\downarrow} +
|\psi(d^{\downarrow})|^2) \right].
\ee

\begin{table}
\begin{center}
{\scriptsize
\begin{tabular}{|c|c|c|c|c|c|c|c|c|c|c|c|c|}       \hline
 & & \multicolumn{5}{c|} {Without configuration mixing} &
\multicolumn{6}{c|} {With configuration mixing}\\ \cline{3-13} 
Para- & Expt. & CQM & \multicolumn{2}{c|} {$\chi$QM}  &
\multicolumn{2}{c|} {$\chi$QM}
& $\phi$ & CQM$_{gcm}$ & \multicolumn{2}{c|} {$\chi$QM$_{gcm}$}
& \multicolumn{2}{c|} {$\chi$QM$_{gcm}$} \\
meter & value&   & \multicolumn{2}{c|} {with SU(3)}  &
\multicolumn{2}{c|} {with SU(3)} & & & \multicolumn{2}{c|}
{with SU(3)}  & \multicolumn{2}{c|} {with SU(3)} \\
 & &   & \multicolumn{2}{c|} {symmetry}  &
\multicolumn{2}{c|} {symmetry} & &
& \multicolumn{2}{c|} {symmetry}  &
\multicolumn{2}{c|} {symmetry} \\  
&&&\multicolumn{2}{c|}{}& \multicolumn{2}{c|} {breaking} &&&
\multicolumn{2}{c|}{}& \multicolumn{2}{c|} {breaking} \\ \hline
& & & NMC  & E866  & NMC &E866 & & & NMC & E866 & NMC & E866 \\
\cline{4-7} \cline{10-13}
 & & & & & & & 20$^o$ & 1.26 &
 .74 & .76 & .90 & .92  \\
 $\Delta$ u & 0.85 $\pm$ 0.05  & 1.33 & .79 & .81 & .96 & .99 &
  18$^o$ & 1.27 &
.75 &.77 & .91 & .93  \\
& {\cite{adams}} & & & & & & 16$^o$ & 1.28 &
.76 & .78 & .92 & .94 \\    \hline

 & & & & & & & 20$^o$ &
 -0.26 & -0.30 & -0.31  & -0.32 & -0.34 \\
 $\Delta$ d & -0.41  $\pm$ 0.05  & -0.33 & -0.35 & -0.37 & -0.40 &
  -0.41 & 18$^o$ & -0.27 &-0.31 & -0.32 & -0.33 & -0.35 \\
   & {\cite{adams}} & & & & & & 16$^o$ &
  -0.28 & -0.32 &-0.33 & -0.34 & -0.36 \\     \hline

&&&&&&&&&&&& \\
$\Delta$ s &-0.07  $\pm$ 0.05 & 0 & -0.1 & -0.12 & -0.02 &
-0.02 & & 0 & -0.1 & -0.12 & -0.02 &-0.02 \\  
& {\cite{adams}} &&&&&&&&&&& \\ \hline

 & & & & & & & 20$^o$ &
1.52 & 1.04 & 1.07 & 1.22 & 1.26 \\
$G_A/G_V$ & 1.267  $\pm$ .0035 & 1.66 & 1.14 & 1.18 & 1.35 &
1.40 & 18$^o$ & 1.54 & 1.06 & 1.09 & 1.24 & 1.28  \\
 & {\cite{PDG}}& & & & & & 16$^o$ &         
1.56 & 1.08 & 1.11 & 1.26 & 1.30  \\   \hline

 & & & & & & & 20$^o$ &
1 & .64 & .69 & .62 & .62 \\
$\Delta_8$ & .58  $\pm$ .025 & 1 & .64 & .68 & .60 & .62  &
18$^o$ & 1 & .64 & .69 & .62 & .62  \\
 & {\cite{PDG1}}& & & & & & 16$^o$ &         
1 & .64 & .69 & .62 & .62  \\  \hline

& & & & & & & 20$^o$ & -.26 &
 -.20 & -.19 & -.30 & -.32  \\
 F-D & -.34   & -.33 & -.25 & -.25 & -.38 & -.39 & 18$^o$ 
& -.27 &
-.21 & -.20 & -.31 & -.33  \\
& & & & & & & 16$^o$ & -.28 &
-.22 & -.21 & -.32 & -.34 \\ \hline

& & & & & & & 20$^o$ & .71 &
 .68 & .70 & .61 & .59  \\
 F/D & .575   & .67 & .64 & .65 & .56 & .56 & 18$^o$ 
& .70 &
.67 &.69 & .60 & .58  \\
& & & & & & & 16$^o$ & .69 &
.66 & .68 & .59 & .57 \\ \hline

& & & & & & & 20$^o$ & .63 &
 .42 & .44 & .46 & .47  \\
 F & .462   & .665 & .445 & .465 & .49 & .505 & 18$^o$ 
& .635 &
.425 &.445 & .465 & .475  \\
& & & & & & & 16$^o$ & .64 &
.43 & .45 & .47 & .48 \\ \hline

& & & & & & & 20$^o$ & .89 &
 .62 & .63 & .76 & .79  \\
 D & .794  & 1 & .695 & .715 & .87 & .895 & 18$^o$ 
& .905 &
.635 &.645 & .775 & .805  \\
& & & & & & & 16$^o$ & .920 &
.65 & .66 & .79 & .82 \\   \hline

\end{tabular}}
\caption{The calculated values of
spin polarization functions $\Delta u, ~\Delta d, ~\Delta s$,
and quantities dependent  on these: $G_A/G_V$ and $\Delta_8$ 
both for  NMC and E866 data with the symmetry breaking 
parameters obtained by $\chi^2$ minimization in the $\chi$QM with 
one gluon generated configuration mixing ($\chi$QM$_{gcm}$) and
SU(3) symmetry breaking.}
\end{center}
\end{table}

Using the spin structure from the above Equation
we can calculate the spin polarizations, which come out to be

\[  \Delta u ={ \rm cos}^2 \phi \left[\frac{4}{3}-\frac{a}{3}
   (7+4 \alpha^2+ \frac{4}{3} \beta^2
   + \frac{8}{3} \zeta^2)\right] \]
\be
   + {\rm sin}^2 \phi \left[\frac{2}{3}-\frac{a}{3} (5+2 \alpha^2+
  \frac{2}{3} \beta^2 + \frac{4}{3} \zeta^2)\right], 
\ee

\clearpage

\[  \Delta d ={\rm cos}^2 \phi \left[-\frac{1}{3}-\frac{a}{3}
  (2-\alpha^2- \frac{1}{3}\beta^2- \frac{2}{3} \zeta^2)\right] \]
\be
  + {\rm sin}^2 \phi  \left[\frac{1}{3}-\frac{a}{3} (4+\alpha^2+
  \frac{1}{3} \beta^2 + \frac{2}{3} \zeta^2)\right],
\ee
   
\be
  \Delta s = {\rm cos}^2 \phi[-a \alpha^2] +
  {\rm sin}^2 \phi[-a \alpha^2]= -a \alpha^2.
\ee

Before we present our results it is perhaps desirable to discuss
certain aspects of the symmetry breaking parameters employed here.
As has been considered by Cheng and Li {\cite{cheng}}, the singlet
octet symmetry breaking parameter
$\zeta$ is related to $\bar u- \bar d$ asymmetry
{\cite{{NMC},{GSR},{E866}}. We have also
taken $\zeta$ to be responsible for the $\bar u-\bar d$ asymmetry
in the $\chi$QM with SU(3) symmetry breaking and configuration
mixing.
Further the parameter $\zeta$ is constrained 
\cite{{NMC},{johan},{E866}}
by the expressions $\zeta=-0.7-\beta/2$ and $\zeta=-\beta/2$
for the NMC and E866 experiments respectively, which essentially
represent the fitting of deviation from Gottfried sum rule
{\cite{GSR}}.

In Table 4, we have presented the results of our calculations
pertaining to spin polarization functions $\Delta u, ~\Delta d,
~\Delta s$ and related parameters including the  hyperon
$\beta$-decay parameters dependent on spin polarizations
functions. The value of the mixing angle is taken to be
$\phi \simeq$ 20$^o$, a value dictated by consideration of
neutron charge radius, as discussed earlier.
In the table, however,  we have considered a few
more values of the mixing parameter $\phi$ in order to study
the variation of spin distribution functions with  $\phi$.
The parameter $a$ is taken to be 0.1, as considered by other 
authors \cite{{song},{johan},{cheng},{eichten}}. 
Further, while presenting the results of  SU(3) symmetry 
breaking case without configuration mixing $(\phi=0^o)$, 
we have used the same values 
of parameters  $\alpha$ and  $\beta$, primarily to understand
the role of configuration mixing for this case. 
The SU(3) symmetry calculations are obtained by taking
$\alpha= \beta=1, \phi=20^o$ and $\alpha= \beta=1, \phi=0^o$
respectively for with and without configuration mixing. For
the sake of completion, we have also presented the results
of CQM with and without configuration mixing.
                                                                                                                            
In order to appreciate the role of
configuration mixing in affecting the fit, we first  compare the
results of CQM with those of CQM$_{gcm}$ {\cite{hd}}.
One observes that
configuration mixing corrects the result of the quantities
in the right direction but this is not to the desirable level.
Further, in order to understand the role of configuration mixing
and SU(3) symmetry with and without breaking  in $\chi$QM,
we can compare the results of $\chi$QM with SU(3)
symmetry to those of $\chi$QM$_{gcm}$ with SU(3)
symmetry. Curiously $\chi$QM$_{gcm}$ compares unfavourably with
$\chi$QM in case of most of the calculated quantities.
This indicates that configuration mixing alone is not enough
to generate an appropriate fit in  $\chi$QM.
However when  $\chi$QM$_{gcm}$ is used with SU(3) and axial
U(1) symmetry breakings then the results show uniform
improvement over the corresponding results of $\chi$QM with
SU(3) and axial U(1) symmetry breakings. To summarize the
discussion of these results, one finds that both configuration
mixing and symmetry breaking are very much needed to fit the
data within $\chi$QM.

\begin{table}
\begin{center}
{\small
\begin{tabular}{|c|c|c|c|c|c|c|}       \hline
              
Parameter & Expt. & CQM & \multicolumn{2}{c|} {$\chi$QM}  &
\multicolumn{2}{c|} {$\chi$QM} \\
 & value&   & \multicolumn{2}{c|} {with SU(3)}  &
\multicolumn{2}{c|} {with SU(3)}  \\
 & &   & \multicolumn{2}{c|} {symmetry}  &
\multicolumn{2}{c|} {symmetry}  \\  
&&&\multicolumn{2}{c|}{}& 
\multicolumn{2}{c|} {breaking} \\ \hline                            
& & & NMC  & E866  & 
NMC &E866                               \\
\cline{4-7}

$\bar d-\bar u$ &.147 $\pm$ .024 {\cite{NMC}} & 0 &  .147  &
.12 & .147 & .12  \\
                & .100 $\pm$ .015 \cite{E866} &&&&& \\

$\bar u/\bar d$ & 0.51 $\pm$ 0.09 {\cite{baldit}} & 1 & .53 &
.63 & .53 &  .63  \\
                & 0.67 $\pm$ 0.06 \cite{E866} &&&&& \\

$I_G$ & .235  $\pm$ .005 {\cite{NMC}}& 0.33 & .235 & .253 &
.235 & .253  \\
      &.266     $\pm$ .005 {\cite{E866}}  &&&&& \\

$\frac{2 \bar s}{\bar u+ \bar d}$ & .477 $\pm$ .051 {\cite{ao}} &
&1.9 & 1.66  &  .62 & .38 \\

$\frac{2 \bar s}{u+d}$ & .099 $\pm$ .009 {\cite{ao}} & 0 & .26 &
 .25 &   .09 & .06  \\

$f_s$ &  .10 $\pm$ 0.06 {\cite{ao}} & 0 & .19 &
.18 &   .07 & .05  \\

$f_3/f_8$ & .21 $\pm$ 0.05 {\cite{cheng}} & .33 & .33 & .33 &
.21 &   .21 \\
\hline

\end{tabular}}
\caption{The calculated values of quark distribution functions
and other dependent quantities as calculated in the $\chi$QM
with and without SU(3) symmetry breaking both for NMC and E866
data, with the same values of symmetry breaking parameters as
used in spin distribution functions and hyperon
$\beta$ decay parameters.}
\end{center}
\end{table}

In order to have a unified fit to spin polarization
functions as well as quark distribution functions, we have
presented in Table 5 the various quark distribution functions
with the symmetry breaking parameters used in the case of
$\chi$QM with symmetry breaking and configuration mixing both
for NMC and E866 data. The general survey of Table 5
immediately makes it clear that the success achieved in the
case of spin polarization functions is very well maintained in
this case also. The calculated values hardly leave anything to
be desired  both for the NMC and E866 data.

We find that $\chi$QM$_{gcm}$ with SU(3) symmetry breaking is
able to give a satisfactory unified fit for spin and quark
distribution functions, with the symmetry breaking parameters
$\alpha=.4$, $\beta=.7$ and the mixing angle
$\phi=20^o$,  both for NMC as well as the most recent E866 data.
In particular, the agreement in the case of  
$G_A/G_V, ~\Delta_8$, F, D, $f_s$ and $f_3/f_8$,  is quite striking. 
It is found that configuration mixing improves the CQM results,
however in the case of  $\chi$QM with SU(3) symmetry the results
become worse. The situation changes completely when SU(3) symmetry
breaking and configuration mixing are included simultaneously. 
Thus, it seems that both configuration mixing as well as symmetry 
breaking are very much needed to fit the data within $\chi$QM.

\section{Summary and conclusion}
In the last 5 decades there has been phenomenal growth in
the understanding of the question: What is inside the nucleon?
The early 60's saw the emergence of unitary symmetry and
consequently the Quark Model. That quark are point like
constituents of nucleon was formally established by deep
inelastic scattering experiments. The emergence of the
Standard Model as an extension of electroweak unification
laid the foundation of the basic tenants of Quantum
Chromodynamics $-$ the theory describing the $q-q$ and $q-\bar q$
interactions inside the hadrons. QCD being non-Abelian in nature
with non-linear interactions between gluons cannot be solved
exactly in all limits, this gives rise to various effective
models describing the inside of the nucleon for different energy
regions. As has been emphasized earlier,  for the low energy
hadronic matrix elements or phenomena involving the surface of
the nucleon, the CQM with QCD inspired spin-spin forces provides a
simplistic but satisfactory description of the data.
Below the surface of the nucleon, in the energy scale
$\Lambda_{QCD}< Q < \Lambda_{\chi QM}$, the effective  degrees
of freedom change from constituent quarks to quarks, $q-\bar q$
pairs, gluons and Goldstone bosons. Mathematically,
in this region the wave function of the nucleon is described
by Equation (\ref{chil}). The wavefunction of nucleon in this
region becomes more complicated as it has to incorporate more
degrees of freedom.
The Deep Inelastic region, with the dynamics of the quarks
described by the QCD Lagrangian  is given by Equation (\ref{disl}).
The deep inside of the nucleon is characterized by
quarks, gluons,  $q-\bar q$ pairs, with hardly any interactions
among themselves.

%%%%%%%%% IF YOU HAVE PS OR EPS FILES FOR FIGURES%%%%%%%%
%\begin{figure}
%\centerline{\epsfxsize=12cm\epsffile{figure1.eps}}
%\caption{Figure caption}
%\end{figure}

\end{document}